\documentclass[]{article}

\usepackage{fullpage}

\usepackage{graphicx}        
\usepackage{multicol}        
\usepackage{multirow}

\usepackage{amsmath}

\usepackage{listings}
\usepackage{paralist}
\usepackage{framed}
\usepackage[boxed]{algorithm2e}
\usepackage{subfigure}
\usepackage{algorithmicx}

\begin{document}
%

\title{A Prediction Model for the Probability of SLA Matching in Consumer Provider Contracting of Web Services}

\author{Werner Mach, Benedikt Pittl and Erich Schikuta\\
  University of Vienna, Faculty of Computer Science \\
  Research Group Workflow Systems and Technology \\
  W\"ahringerstr. 29, A-1090 Vienna, Austria\\
  email: erich.schikuta@univie.ac.at}

\date{}

\maketitle

\begin{abstract}

Future e-business models will rely on electronic contracts which are agreed dynamically and adaptively by web services.
Thus, the automatic negotiation of Service Level Agreements (SLAs) between consumers and providers is key for enabling service-based
value chains.

The process of finding appropriate providers for web services seems to be simple. Consumers contact several providers and take the provider which offers the best matching SLA.
However, currently consumers are not able forecasting the probability of finding a matching provider for their requested SLA. So consumers contact several providers and check if their offers are matching. In case of continuing faults, on the one hand consumers may adapt their Service Level Objects (SLOs) of the required SLA or on the other hand simply accept offered SLAs of the contacted providers.

By forecasting the probability of finding a matching provider, consumers could assess their chances of finding a provider offering the requested SLA. If a low probability is predicted, consumers can immediately adapt their SLOs or increase the numbers of providers to be contacted.

Thus, this paper proposes an analytical forecast model, which allows consumers to get a realistic assessment of the probability to find matching providers. Additionally, we present an optimization algorithm based on the forecast results, which allows adapting the SLO parameter ranges in order to find at least one matching provider.
Not only consumers, but also providers can use this forecast model to predict the prospective demand. So providers are able to assess the number of potential consumers based on their offers too.

Justification of our approach is done by simulation of practical examples checking our theoretical findings.
\end{abstract}




\section{Introduction}

Due to the shift of a capital expenditure (CAPEX) to an operational expenditure (OPEX) focused economy nowadays, the importance of digital marketplaces (aka market spaces) is increasing dramatically. Hereby the paradigm of service oriented utility computing is a pillar of this development for the Internet.

A digital market is the culmination point of stake-holders with integration of services along each link in
a value chain. Services corresponding to different partners are negotiated and contracted in a consumer-producer
manner resulting in added value.

The last two decades of Information Technology (IT) development has witnessed the  specific efforts done to make John McCarthy's vision of ``Resource/Service as a utility'' a reality. Utility computing is providing basics for the current day resource utilization. Cluster, grids and now cloud computing have made this vision a reality. Cluster computing offered more centralized resource pool, while grid computing remained in need via hardware and computation cycles offerings to the scientific community. Grid computing models observed a deadlock after the introduction of cloud computing concepts. This situation was a result of missing economic business models, although some work was done on this issue even by the authors, e.g. business models~\cite{schiki06,schiki07,schiki09} and enabling security issues~\cite{schiki05}. Based on Pay-as-you-use criteria, cloud computing is still in early stage. However, the economic component of cloud computing is a central focus of our actual research~\cite{MachPittlSchikuta13,cbi13}.

Cloud computing by its characteristic of metered service by negotiable Service Level Agreements (SLAs) paves the way to the realization of these digital markets. Thus, in our previous work~\cite{MachSchikuta2012} we introduced a negotiation framework for consumer provider contracting of web services. This framework comprises a consumer agent as well as a provider agent both implementing a negotiation and re-negotiation engine as well as a knowledge base containing business rules which are executed during negotiation~\cite{MachPittlSchikuta13}. The goal of the agents in this framework is to negotiate service parameters, like storage, maximum down time, upload and download bandwidth~\cite{Buyya2011} between consumer and provider and consequently agree to a web service via SLA.
A SLA consists of several so called Service Level Objectives (SLO). A SLO details one performance characteristic of a SLA. Therefore SLOs determine the Quality of Service (QoS)~\cite{wieder2010,schiki14}. In this paper we use the terms SLO parameters and service parameters synonymously.

We strongly believe that SLAs are the key mechanism for enabling market value chains and new business models for Cloud \cite{schiki04,schiki10,schiki12}.
To fully exploit the potential of open markets, a large number of providers and
consumers is necessary. The large number of potential traders might inflate
the variety of resources which leads to the problem that the supply and
the demand are spread across a wide range of resources. To avoid restrictions
for traders, an approach enabling them to define their resources (or requirements)
freely while facilitating SLA matching is needed. Current adaptive SLA
matching mechanisms are based on semantic ontologies like the Web Ontology
Language (OWL)~\cite{wsagreement2007} and OWL-S (former DAML-S) and other semantic technologies.
However, none of these approaches addresses the issues of the open
market and deals with (semi-) automatic definitions of SLA mappings enabling
negotiations between inconsistent SLA templates.

%

Before starting negotiation the consumer agent needs to find an appropriate provider. 
Obviously the consumer is interested in answers to the following questions:

\begin{itemize}
	\item What is the probability of finding a provider offering a SLA matching my request?
	\item Which service parameters of the SLA, and to what extent, have to be adapted in order to find at least one matching provider?
	\item How many providers should I contact in order to be practically sure to find at least one matching provider?
\end{itemize}


Within this paper we try to give answers to these questions by defining an analytical model forecasting the probability of finding appropriate providers for a requested SLA.
We follow a twofold approach by defining a theoretical model and proving its applicability by practical simulation.
Further we present an algorithm optimizing a consumer's SLA in order to practically ensure finding at least one appropriate provider. To omit redundancy, we focus our analysis on the consumer's point of view. However, all answers given can be applied to the provider as well.
The justification of our proposed approach is done within a simulation framework proving our theoretical findings on practical examples.

The layout of the paper is as follows. In the next section~\ref{sec:overlap} we present the assumptions of our analytical forecast model and identify the distribution function for SLO interval overlaps by a simulation approach. The analytical forecast model is presented in section~\ref{sec:calculation} followed by the negotiation interval optimization algorithm in section~\ref{sec:optimization}. To elaborate practically the analytical method a comprehensive use case and its justification by simulation is given in section~\ref{sec:usecase}. Finally, in section~\ref{sec:prototype}, the simulation prototype implementing our negotiation model is described. The paper is closed by a summary in section~\ref{sec:conclusion}.

\section{Overlap Distribution Function}
\label{sec:overlap}

Our forecast model is based on some assumptions regarding providers and consumers.
\begin{itemize}
	\item \textbf{Provider}
	
	 The SLO parameters offered by the providers are ranges and can consequently be represented as intervals. We call such an interval \textit{provider interval}. Some of the intervals may cover the whole market whereas others may be very small. The size and position of the ranges of the providers are independent of each other.
	
	\item \textbf{Consumer}
	
	The consumers' requested SLO parameters are ranges and can be represented as intervals, too. Such an interval is called \textit{consumer interval}.
\end{itemize}

\begin{figure}[htb]
\centering
	\includegraphics[width=0.5\linewidth]{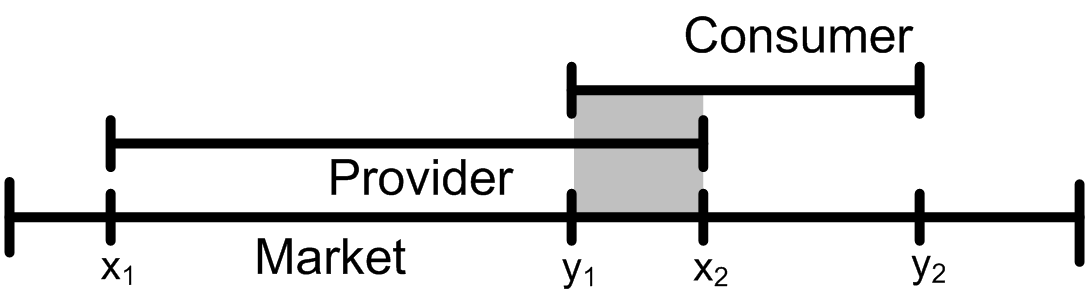}
	\caption{Provider and consumer interval}
\label{fig:intervals}
\end{figure}

Figure ~\ref{fig:intervals} illustrates a typical situation on the market: A provider offers a SLO parameter from $x_{1}$ to $x_{2}$. The consumer requests this SLO parameter from $y_{1}$ to $y_{2}$. The overlapping range for this SLO parameter reaches from $y_{1}$ to $x_{2}$. We call the difference $x_{2} - y_{1}$ negotiating range which will be relevant for the negotiation between provider and consumer.

As shown in figure ~\ref{fig:intervals}, the probability of finding a provider offering a service parameter which matches the consumer's request is equal to the probability that these two intervals are overlapping.

For forecasting we need both the requested SLO parameters ranges of the consumer and the offered service parameter ranges of the provider. Unfortunately, the service parameter ranges of the providers can change quickly and no repository containing all this data in real-time is available in practice. Therefore we are forced to make assumptions about the provider ranges: In our simulation, which is described in section ~\ref{sec:simulation}, the length as well as the position of the \textit{provider intervals} are generated randomly. 
Thus, the consumer's SLO parameter ranges are covered by the providers with the same probability, independently of the ranges' position and the SLO parameter type.

Some work on overlapping intervals was done in the past and is condensed at \textit{mathpages.com}~\cite{Mathpages2013}. On this website a mathematical analysis for the probability calculation of overlapping, randomly generated, intervals is derived. However, this approach has one essential constraint: it is assumed that the intervals have the exact same length. Regarding figure ~\ref{fig:intervals}, this constraint would enforce that the consumer's and the provider's range have the same length. Obviously, this constraint violates our assumption that the provider's interval length is created randomly and may be different to the consumer's interval length.
Thus, to the best of the authors' knowledge, no analysis for calculating the probability of overlapping intervals with different lengths exists.

One goal of our approach is to keep the forecast model easy adaptable regarding to changes of interval generation and modifications of interval lengths.
This leads to build a simulation which generates \textit{provider intervals} and checks if they overlap with \textit{consumer intervals}. Based on the simulation's result it is possible to determine the probability distribution of overlapping intervals with different lengths.
If the interval generation changes or the length limits of intervals are modified the simulation adapts respectively. In turn the modified simulation will be executed several times in order to determine the new probability distribution and consequently delivering new forecasts. The calculation details will be explained in the following sections.

\subsection{Probability Function Identification}
\label{sec:simulation}
We applied a simulation approach for identifying the interval overlapping probability function which is the basis for our analytical model in section~\ref{sec:calculation}. Goal of the
 simulation is to generate intervals and check if providers' offers matches consumers' requests. By executing this simulation with a large number of experiments, we calculate the underlying probability distribution of \textit{provider and consumer interval} overlaps.

Algorithm~\ref{alg:simulation} sketches execution of the simulation.

\begin{algorithm}
  \KwData{SLA from consumer}
  \KwResult{Number of matching providers}
  \For {each experiment}{
  	\For {each provider on market}{
  	 	\For {each service parameter S}{
   				generate random interval representing provider range\;
    		  \If {consumer and \textit{provider interval} $\lnot$overlapping} {
    		  	/*provider $\lnot$matches*/
   		 	  	
    		     break\;
   	 	  }
    		  \If {S is last parameter in SLA} {
    		     /*provider matches*/
    		
    		     start next experiment
    		  }
 		 }
	 }
  }
\caption{Finding matching providers}
\label{alg:simulation}
\end{algorithm}

The algorithm counts the number of providers until a matching provider is found. By executing the algorithm with a large number of experiments, we get a distribution which shows how many providers are necessary to find a match. Our results and a more detailed description will be presented in section~\ref{subsec:simresult}.


A key issue within the simulation is the interval generation representing the provider range. The \textit{consumer interval's} length and position are given. 
We assume that the real market encompasses many providers offering small intervals and few providers offering big intervals. This relation should be represented in our simulation. The position of the \textit{provider intervals} has to be set so that the \textit{consumer interval's} position is negligible.

\textit{Provider intervals} are generated in two steps.

\begin{enumerate}
	\item Without loss of generality we make two constraints:
	
	\textsf{\textbf{Constraint 1:} The whole market of each SLO parameter has a minimum of $0$ and a maximum of $100$}.
	
	\textsf{\textbf{Constraint 2:} The length of the \textit{provider interval} is between $0$ and $100$.}
	
	Each \textit{provider} and \textit{consumer interval} lies within this market range. A provider can offer a service parameter covering the whole market or just $1$ unit of the market. The length of the \textit{provider interval} is generated based on the uniform distribution considering the minimum and maximum of Constraint $2$.
	Calculating the distribution of many small intervals and few long intervals, a further step is necessary.
	
	\item After calculating the temporary uniform distributed length, the position of the intervals has to be generated. Simply randomly selecting a start point between $0$ and $100-$\textit{provider interval length} would lead to a unsatisfactory result: A \textit{consumer interval} placed in the near of the middle will have a higher probability of matching with a provider than those placed near of margins. The following two examples clarify this property.
\begin{inparaenum}[(i)]
    \item A \emph{consumer interval}, placed \textbf{leftmost}, matches with \textit{provider intervals} whose start points are within the \textit{consumer interval}.
    \item A \textit{consumer interval}, placed \textbf{at the middle}, matches with \textit{provider intervals} whose start points are within the \textit{consumer interval}. Further, a \textit{provider interval} whose start point is smaller than the \textit{consumer interval's} start point but whose end point is greater than the \textit{consumer interval's} start point matches with the \textit{consumer interval}.
\end{inparaenum}	

However, as already mentioned, the position of the \textit{consumer interval} should be negligible. To avoid this problem, we determine the position of the interval by defining flexible boundaries as shown by following equation.
	\begin{equation}
		\begin{aligned}
		& min=0-length/2 \\		
		& max=100+length/2 \\		
		& center_{\text{interval}}=random(min, max)		
		 \end{aligned}
		\label{equ:randomBoundaries}
	\end{equation}

	With this position generation the \textit{consumer interval's} position is irrelevant. 
	The so generated random intervals are \textit{provider intervals}.
	 However, this position calculation can lead to \textit{provider intervals} adapting the market range as figure~\ref{fig:intervalsExceeded} shows.
	
	\begin{figure}[htb]
	\centering
		\includegraphics[width=0.8\linewidth]{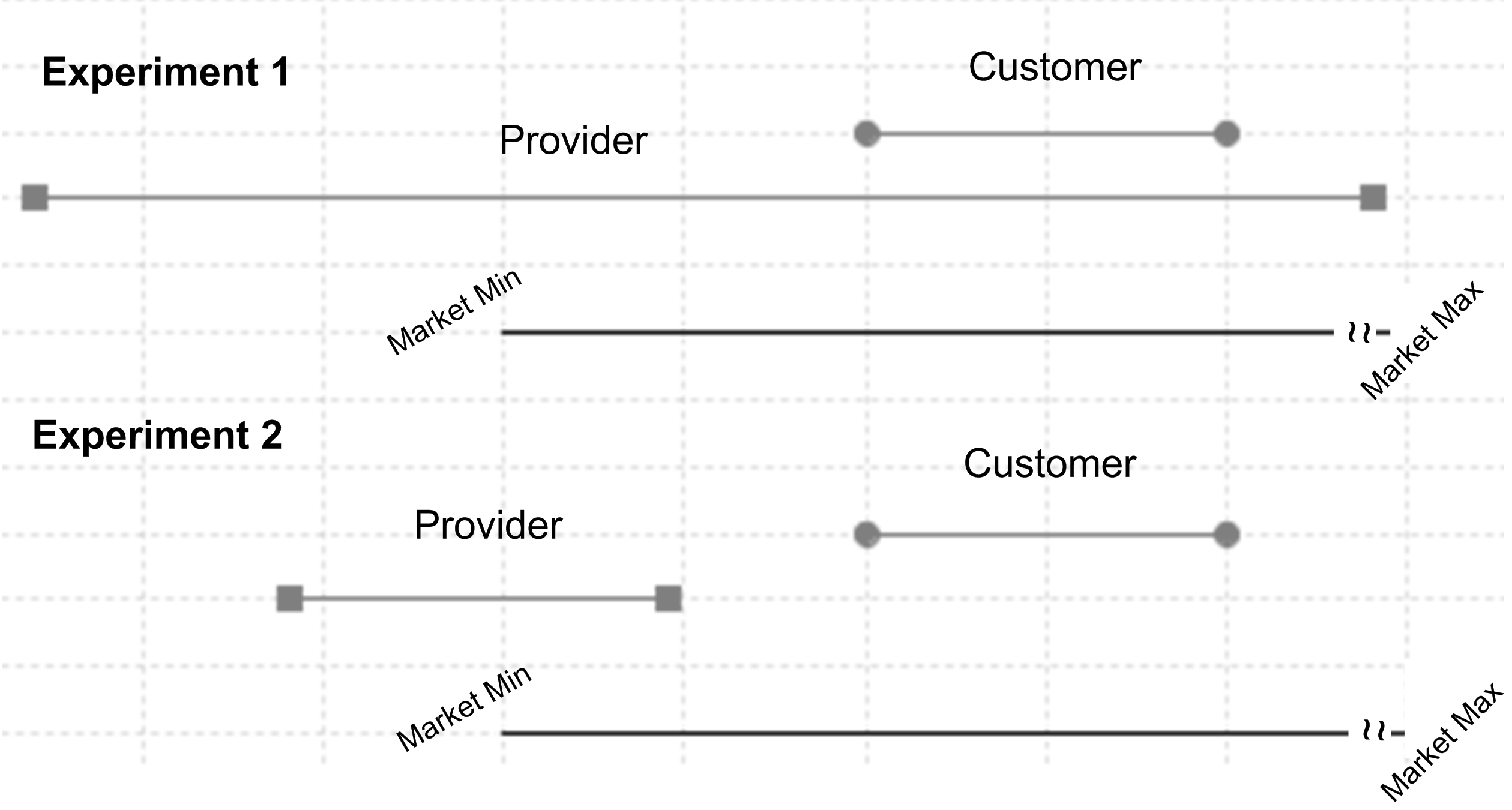}
		\caption{Provider and consumer interval}
	\label{fig:intervalsExceeded}
	\end{figure}	
	
	As the \textit{consumer interval} is in the market range, we can simply cut the protruding interval part. This cut leads to few providers with a big range and a lot of providers with a small range. The histogram in figure~\ref{fig:providerIntervalLengthHistogram} represents exemplary this distribution.
	
	\begin{figure}[htb]
	\centering
		\includegraphics[width=0.9\linewidth]{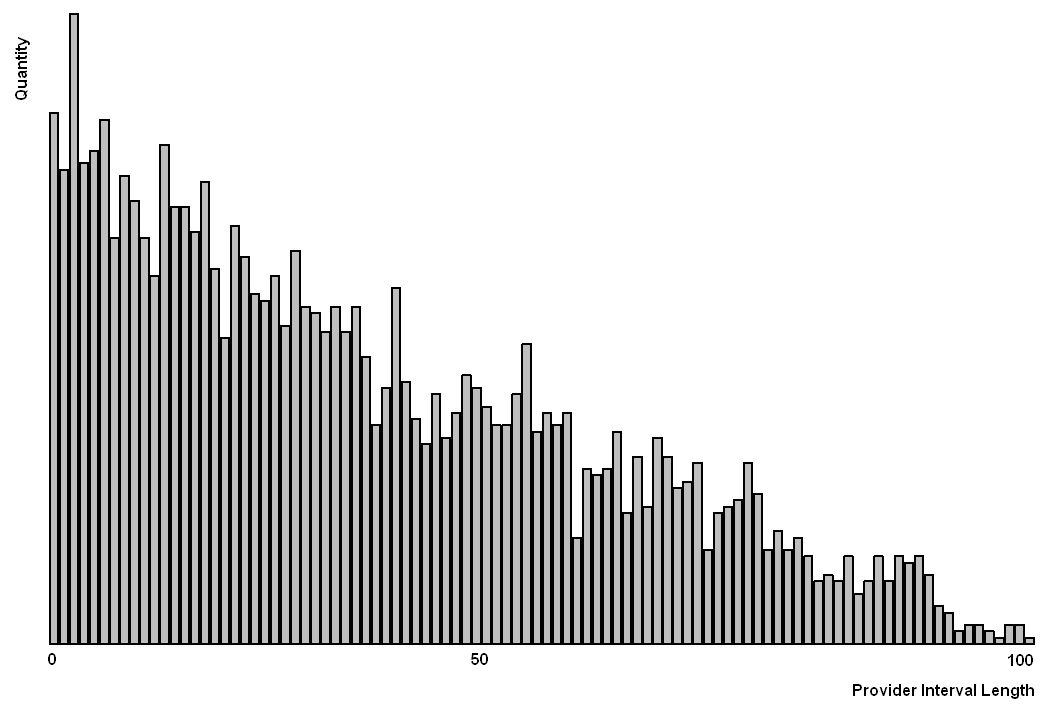}
		\caption{Provider interval length simulation}
	\label{fig:providerIntervalLengthHistogram}
	\end{figure}	
	
\end{enumerate}

\subsection{Simulation Result}
\label{subsec:simresult}
The first simulation coped with consumer SLAs which contain one SLO parameter only. The goal was to count the number of providers needed to find one provider offering the requested SLO parameter.
We executed the simulation with 1 million experiments for each \textit{consumer interval} with a length of $10$, $20$, $30$, $40$, $50$, $60$, $70$, $80$, $90$ and $100$.
Table~\ref{tab:distributionSimulationResult} shows the results of the simulation for a \textit{consumer interval} with a length of $20$. In $448844$ cases the first provider was the first matching provider, in $247577$ cases the second provider was the first matching provider and so on.

\begin{table}
\caption{Simulation results for interval length 20}
\begin{center}
\begin{tabular}{|l|c|c|c|c|c|}
\hline
\hline
provider & 1 & 2 & 3 & 4 & \dots  \\
first matches & 448844 & 247577 & 136070 & 75437 & \dots \\
\hline
\end{tabular}
\end{center}
\label{tab:distributionSimulationResult}
\end{table}

The total number of experiments was 1000000.
%
%
%

\begin{figure}[htb]
	\centering
		\includegraphics[width=0.9\linewidth]{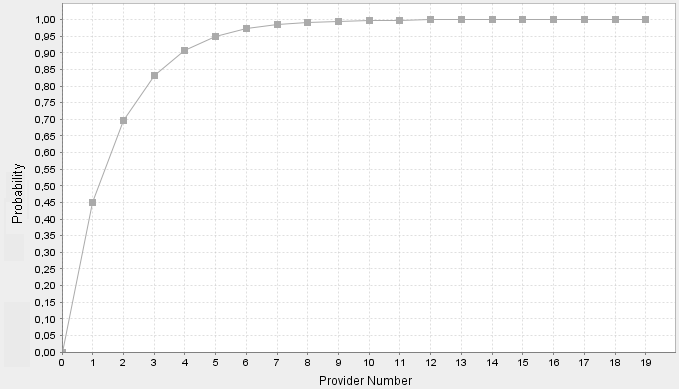}
		\caption{Probability of finding matching providers for 1 interval of length $20$}
	\label{fig:probabilitySimulationScreenshot}
	\end{figure}

Figure~\ref{fig:probabilitySimulationScreenshot} visualizes the results of the simulation represented in table~\ref{tab:distributionSimulationResult}. 
The abscissa shows the provider's number, the ordinate shows the cumulative numbers of first matches, expressed as percentage. As the simulation shows, about $70\%$ of first matches occur in the first two providers.

The simulation reveals, that in $44,9\%$ of all experiments, the \textit{consumer interval} overlaps with the first randomly generated \textit{provider interval}.
Based on this simulation result, we conclude that the probability of overlapping of a \textit{consumer interval} of length $20$ and a random \textit{provider interval} is about $44,9\%$.

The results shown in table~\ref{tab:distributionSimulationResult} are valid for a \textit{consumer interval} with a length of $20$.  Table~\ref{tab:distributionSimulationResultComprehensive} represents the numbers of first matches of the first provider for all the other \textit{consumer interval} lengths expressed in percentages. Again, we conclude that these numbers represent the probabilities of overlapping with a random generated \textit{provider interval}. With an increasing \textit{consumer interval} length the probability, that the provider matches, increases.

\begin{table}[htbp]
\caption{Simulation results for all interval lengths}
\begin{center}
\begin{tabular}{|l|c|c|c|c|c|}
\hline
\hline
interval length & 10 & 20 & 30 & 40 & 50  \\
first matches & 38,1\% & 44,9\% & 51,7\% & 58,7\% & 65,6\% \\
\hline
interval length  & 60 & 70 & 80 & 90 & 100  \\
first matches  & 72,4\% & 79,3\% & 86,2\% & 93,2\% & 100\%   \\
\hline
\end{tabular}
\end{center}
\label{tab:distributionSimulationResultComprehensive}
\end{table}

The number of experiments which match with the second, third,\dots providers are not relevant for our conclusions.

In our second simulation we simulated the search for a provider offering a matching SLA containing more than one service parameter. If a SLA consists of $n$ service parameters, a matching provider has to offer $n$ service parameters respectively.

\begin{figure}[htb]
	\centering
		\includegraphics[width=0.8\linewidth]{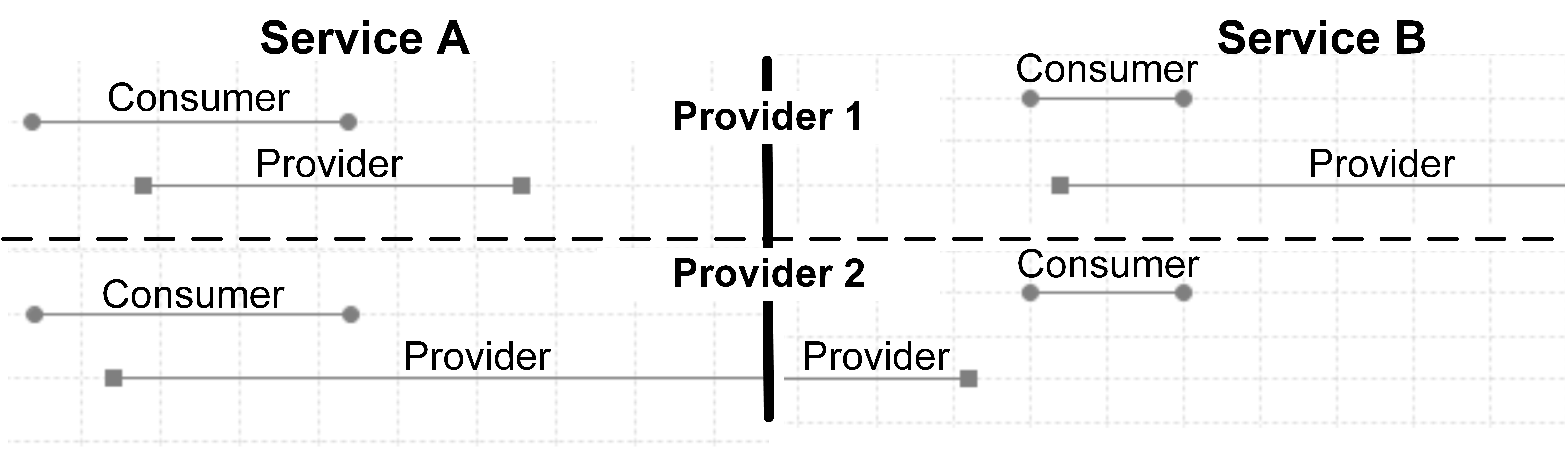}
		\caption{Intervals of 2 providers and 2 services}
	\label{fig:probabilitySimulationSLAScreenshot}
	\end{figure}

Figure~\ref{fig:probabilitySimulationSLAScreenshot} shows an example of finding a SLA containing two service parameters $A,B$. The consumer's service parameter $A$ has the length of $20$ whereas parameter $B$ has a length of $10$. Provider $1$ offers a matching service parameter $A$ as well as a matching service parameter $B$. Consequently provider $1$ offers a matching SLA. Provider $2$ offers a matching service parameter $A$ but service parameter $B$ is not matching. So provider $2$ does not offer a matching SLA.

\begin{figure}[htb]
	\centering
		\includegraphics[width=0.9\linewidth]{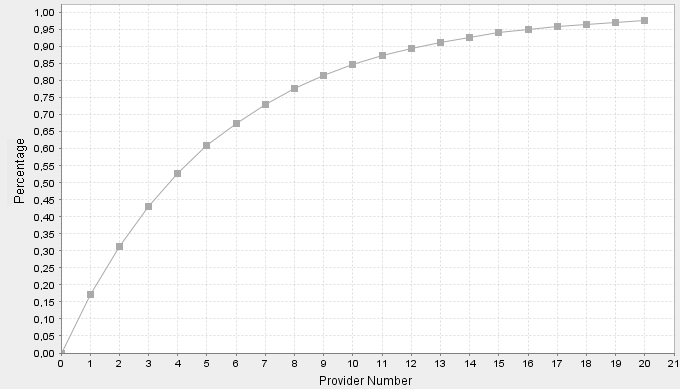}
		\caption{Probability of finding matching providers for 2 intervals of length $10$ and $20$}
	\label{fig:slaProbabilitySimulation}
	\end{figure}

The result of the simulation with these two services A and B is visualized in figure~\ref{fig:slaProbabilitySimulation}. Analogously to figure~\ref{fig:probabilitySimulationScreenshot} this diagram represents the cumulative number of first matches expressed as percentages. The abscissa shows the providers' number whereas the ordinate represents the cumulative number of first matches expressed as percentage. The diagram shows, that about $17\%$ of the first matches occur in the first provider.

Finding a provider offering a matching SLA with more than one service is more difficult than finding a provider with one matching service parameter.

\section{Analytical Prediction Model}
\label{sec:calculation}
\begin{figure}[htb]
	\centering
		\subfigure[Scatter Plot - One Parameter Matching]{\label{fig:sub:a}\includegraphics[width=.4\linewidth]{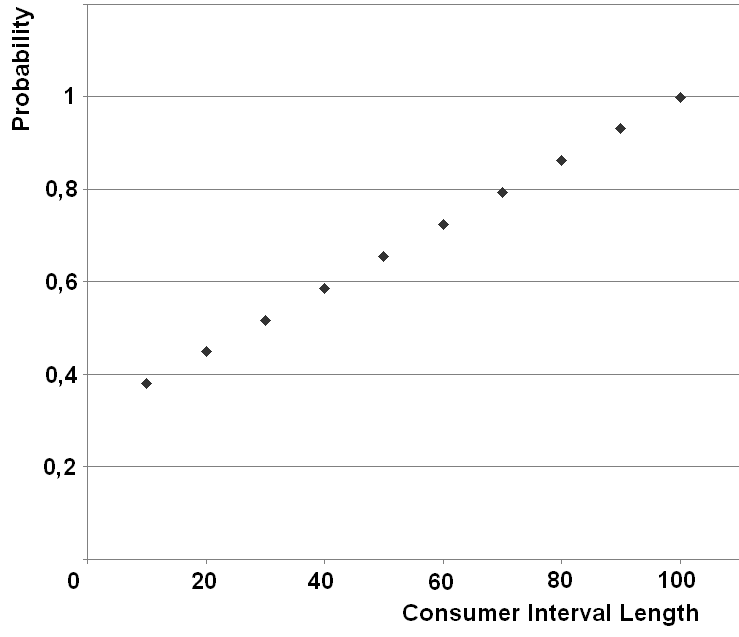}}\hfill
		\subfigure[Trend Line - One Parameter Matching]{\label{fig:sub:b}\includegraphics[width=.4\linewidth]{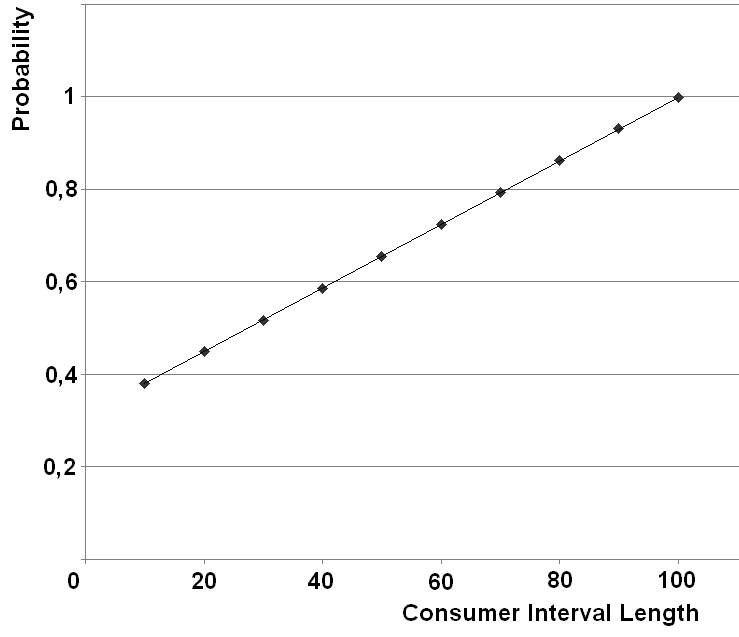}}
		\caption{Overlapping probability for arbitrary length}
	\label{fig:punktWolkeLineareRegression}
	\end{figure}

Based on the simulation results we set up an analytical model forecasting the probability of finding matching SLAs given
\begin{inparaenum}[(i)]
	\item the number of \textit{consumer intervals}
	\item and the length of these \textit{consumer intervals}. 
\end{inparaenum}
First, we examine the simulation results shown in table~\ref{tab:distributionSimulationResultComprehensive}. These numbers represent the overlapping probability of \textit{consumer interval} and \textit{provider interval}. For the forecast calculation, we need a formula returning the probability of overlaps for a \textit{consumer interval} of any length. The results of table~\ref{tab:distributionSimulationResultComprehensive} are visualized in figure~\ref{fig:punktWolkeLineareRegression}. The abscissa represents the \textit{consumer interval} length. The ordinate represents the probability that the \textit{consumer interval} overlaps with a randomly generated \textit{provider interval}. This representation reveals a linear correlation between the \textit{consumer interval} length and the probability that the \textit{consumer interval} overlaps the \textit{provider interval}. To prove the interval-length/probability correlation we use linear regression analysis.

\subsection{Linear Regression Analysis}
\label{subsec:linearRegression}
 The simulation returned results for intervals with lengths $10$, $20$, $30$, $40$, $50$, $60$, $70$, $80$, $90$, $100$. In order to get probabilities for all interval lengths we follow a regression trend line approach.


Precondition for using regression analysis is the classification of the used variables into dependent and independent variables~\cite{archdeacon1994}.
The independent variable is $x$, the interval length, and the dependent variable is $y$, the matching probability.
The depend values are represented as the \textit{first matches} lines in table~\ref{tab:distributionSimulationResultComprehensive}. Interval length are listed in the \textit{interval length} lines in table~\ref{tab:distributionSimulationResultComprehensive}.

First is to calculate the average values of the two variables.

\begin {equation}
\bar{X}=\frac{1}{n} \sum\limits_{i=1}^n x_{i}=55
\end{equation}

\begin {equation}
\bar{Y}=\frac{1}{n} \sum\limits_{i=1}^n y_{i}=0,6901=69,01\%
\end{equation}

The slope of the regression trend line is calculated by
\begin{equation}
s=\frac{\sum\limits_{i=1}^n ((x_{i}-\bar{X})
{(y_{i}-\bar{Y})})}{\sum\limits_{i=1}^n (x_{i}-\bar{X})^{2}}=  0,00688667 \approx 0,0069
\end{equation}

The slope is about $0,69$. The general linear equation is given by 

\begin{equation}
	y-\bar{Y}=s \cdot (x-\bar{X})
\end{equation}

We already calculated the variables $\bar{Y}$, $\bar{X}$ and $s$. Transforming the equation gives the linear expression shown in (\ref{equ:linearRegressionFormular}). This trend line (see figure~\ref{fig:punktWolkeLineareRegression}) describes the predicted probability of overlapping intervals depending on the length of the \textit{consumer interval} $x$.

\begin{equation}	
	y-0,6901=0,00688667 \cdot (x-55)
\end{equation}
\begin{equation}
	y=0,00688667 \cdot x + 0,31133315
	 \label{equ:linearRegressionFormular}
\end{equation}

After calculating the trend line the quality of approximation has to be checked.
The vertical difference between the observed value and the predicted value, based on the trend line, is the residual $R$. 
The sum of all positive and negative residuals has to be $0$~\cite{archdeacon1994}.

\begin{equation}
	R=\sum\limits_{i=1}^n y_{i_{observed}} - y_{i_{predicted}}=0
\end{equation}

The sum of all squared differences ''sum of square error ($SSE$)'' measures the quality of the trend line. Generally, the smaller it is, the better is the approximation by the trend line~\cite{bajpai2009}.


\begin{equation}
	SSE=\sum\limits_{i=1}^n (y_{i_{observed}} -
	y_{i_{predicted}})^2= 2,9333 \cdot 10^{-6}
\end{equation}

The coefficient of determination $R^2$ represents the relation of $SSE$ to $SS_{t}$. $SS_{t}$ represents the total sum of the squared differences from the observed values to the average. By dividing $SSE$ by $SS_{t}$, the $SSE$ is standardized~\cite{moy2000}.

\begin{equation}
	\begin{split}	
	R^2=1-\frac{SSE}{SS_{t}}=1-\frac{\sum\limits_{i=1}^n (y_{i_{observed}} - 	 y_{i_{predicted}})^2}{\sum\limits_{i=1}^n (y_{i_{observed}} - \bar{Y_{i}})^2} \\
	=\frac{\sigma_{y_{predicted}}^2}{\sigma_{y_{observed}}^2}
	\end{split}
\end{equation}
The $R^2$ is a standardized ratio, measuring the goodness of fit with a maximum of 1 and a minimum of 0.  A $R^2$ of 1 means 100\% of the variance is shared between the two variables~\cite{black2011}.
To calculate $R^2$ we need the variances of observed and predicted values.
\begin{equation}
	\sigma_{y_{observed}}^2=\frac{\sum\limits_{i=1}^n (y_{i_{observed}}-
	\bar{Y})^2}{\text{number of values}} = 0,03912689
\end{equation}

\begin{equation}
	\sigma_{y_{predicted}}^2=\frac{\sum\limits_{i=1}^n (y_{i_{predicted}}-
	\bar{Y})^2}{\text{number of values}} = 0,039126635
\end{equation}

$R^2$ represents the ratio of the variances which is in our calculation below nearly 1. This means that the the trend line calculated by equation (\ref{equ:linearRegressionFormular}) is a very good approximation, as it shows a linear relationship between the two variables.

\begin{equation}
	R^2=\frac{\sigma_{y_{predicted}}^2}{\sigma_{y_{observed}}^2} = 0,99999347
\end{equation}

\subsection{Probability Calculation}
Equation (\ref{equ:linearRegressionFormular}) allows to calculate the probability of finding a provider, offering a single matching service parameter depending on the \textit{consumer interval's} length. We call this probability ''single probability''.
In order to get the probability of finding a whole SLA containing more than one SLO all its single probabilities $P_{i}$ have to be combined as shown in equation (\ref{equ:probabilityMultiplicationRule}). It is assumed that the single probabilities are independent of each other. 

\begin{equation}	
	P_{total}=\prod \limits_{i=1}^{n} P_{i} =P_{1} \cdot P_{2} \dots P_{n} \
	\label{equ:probabilityMultiplicationRule}
\end{equation}



\subsubsection{Example 1}
\label{subsubsec:example1}
Given is a consumer requested SLA consisting of three service parameters $A$, $B$ and $C$. Service $A's$ length is $20$, service $B's$ length is $30$ and service $C's$ length is $10$. First, we can calculate the probability of finding a matching provider for each service (equation (\ref{equ:linearRegressionFormular})) by equation (\ref{equ:singleProbabilityCalculation}).

\begin{equation}
	\begin{aligned}
	&P(\text{Service A})=0,006888667 \cdot 20 + 0,31133315= 44,91\% \\		
	&P(\text{Service B})=0,006888667 \cdot 30 + 0,31133315= 51,80\% \\
	&P(\text{Service C})=0,006888667 \cdot 10 + 0,31133315= 38,02\%
	\end{aligned}
	\label{equ:singleProbabilityCalculation}
\end{equation}

The probability of finding a provider, offering a matching SLA, can be calculated by multiplying the single probabilities:

\begin{equation}
	\begin{aligned}
	&P(SLA)=\\
	&P(\text{Service A}) \cdot P(\text{Service B}) \cdot P(\text{Service C})=\\
	&44,91\% \cdot 51,80\% \cdot 38,02\%=8,84\%
	\end{aligned}
\end{equation}

Thus, the probability of finding a matching $SLA$, consisting of the three service parameters $A,$ $B,$ and $C$, is $8,84\%$.

However, for the consumer it is more important to know the probability of finding at least one matching provider by a given number of providers than to know the probability of finding a provider with a matching $SLA$. 

The binomial distribution is a discrete probability distribution. It returns the probability of finding exactly a given number of successes by a given number of experiments. For calculating the binomial distribution the following information is necessary:
\begin{itemize}
	\item the probability of success for one experiment,
	\item the number of experiments, and
	\item the number of the successes
\end{itemize}

In our case, the number of experiments represents the number of providers which are contacted. This factor is limited by the number of providers on market. The number of successes is the number of matching providers. The probability of success for one experiment depends on the consumer's interval length (equation (\ref{equ:probabilityMultiplicationRule})).
The binomial distribution expects that the experiments are independent of each other. This means that finding a matching provider does not influence the result of the other experiments.

The general form for calculating the binomial distribution is given in equation (\ref{equ:binomialDistributionFormula}), where $n$ represents the number of experiments, $k$ the number of successes, and $P(k|p,n)$ represents the probability that exactly $k$ successes occur within $n$ experiments. $p$ is the probability that one event succeeds.

\begin{equation}	
	\begin{aligned}
	&P(k|p,n)=\left( {{n}\atop {k}}\right) \cdot p^{k} \cdot (1-p)^{n-k} \\
	&\left({{n}\atop {k}}\right)=\frac{n!}{(n-k)! \cdot k!}
	\end{aligned}
	\label{equ:binomialDistributionFormula}	
\end{equation}

However, we do not need the probability that exactly $k$ successes occur by executing $n$ experiments as shown in equation (\ref{equ:binomialDistributionFormula}). As already mentioned, we need the probability that at least one success, which represents the finding of one matching provider, occurs. We need to use the inverse probability~\cite{roe2001}.
Thus, we subtract the probability of \textit{no event succeeds} from $1$ in order to get the probability of \textit{at least one event succeeds}:

\begin{equation}
	P(\text{at least one success})=1-P(\text{no success})
	\label{equ:inverseProability}
\end{equation}

\subsubsection{Example 2}
\label{subsec:exampleII}
We extend our example from section~\ref{subsubsec:example1}. The probability of finding a matching SLA is $8,84\%$. The consumer is interested in the probability of finding at least one matching provider. The market, which is consulted by the consumer, has $20$ providers.

First, we calculate the probability of \emph{no success} using equation (\ref{equ:binomialDistributionFormula}).

\begin{equation}
	\begin{aligned}
	&\left({{n}\atop {k}}\right)=\frac{n!}{(n-k)! \cdot k!}=\frac{20!}{(20-0)! \cdot 0!}=1 \\
	&P(k|p,n)=\left({{n}\atop {k}}\right) \cdot p^{k} \cdot (1-p)^{n-k}=\\
	&1 \cdot 0,0884^{0} \cdot (1-0,0884)^{20-0}=0,1571
	\end{aligned}
\end{equation}

The probability of finding exactly zero matching providers, by contacting $20$ providers, is $15,71\%$. Now we have to apply the inverse probability in order to get the probability of finding at least one matching provider.

\begin{equation}
	\begin{aligned}
	&P(\text{at least one success})=\\
	&1-P(\text{no success})=1-0,1571=0,8429
	\end{aligned}
\end{equation}

The probability of finding at least one provider consulting $20$ providers is about $84,29\%$.

\begin{figure}[htb]
	\centering
		\includegraphics[width=0.9\linewidth]{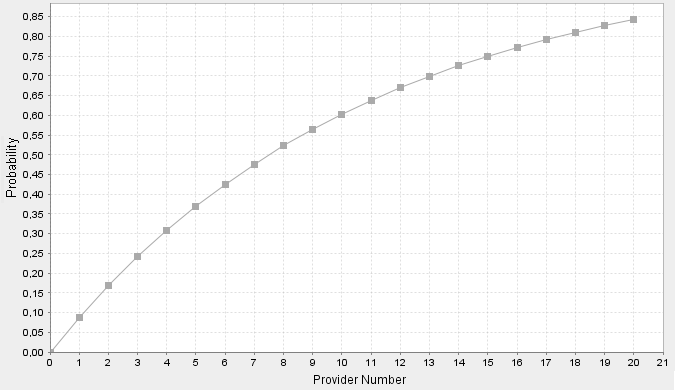}
		\caption{SLA simulation for 20 providers}
	\label{fig:slaSimulationExample}
	\end{figure}

We validate our calculation by comparing the results to the simulation results. Therefore we execute the simulation with a SLA containing the same services as described in the example presented in section~\ref{subsubsec:example1}. Figure~\ref{fig:slaSimulationExample} depicts the simulation result. The probability of finding at least on provider offering a matching SLA is about $84\%$ by consulting $20$ providers.

\begin{figure}[htb]
	\centering
		\includegraphics[width=0.9\linewidth]{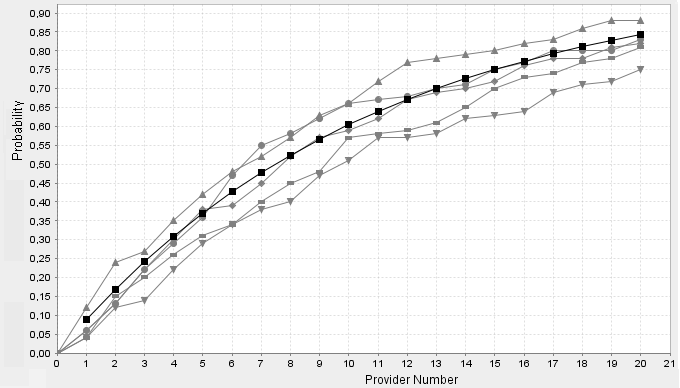}
		\caption{SLA simulation and forecast calculation}
	\label{fig:slaSimulationExampleII}
	\end{figure}

So the simulation delivers approximately the same result as the calculation. This fact is summarized in figure~\ref{fig:slaSimulationExampleII} which illustrates $5$ simulations (grey lines), whereby each simulation was executed with $100$ experiments. The dark line is the calculated forecast. The calculated line is approximately the average of the simulated lines.

Summing up, the following list recaps the steps for calculating the probability to find at least one matching provider:

\begin{itemize}
	\item Calculate the probability of finding a provider for each single service parameter of a SLA using the trend line equation.
	\item Combine the single probabilities by multiplication resulting the probability of finding a provider offering a matching SLA.
	\item Finally, use the binomial distribution in order to calculate the probability of finding at least 1 matching provider.
\end{itemize}

\section{Optimization Algorithm}
\label{sec:optimization}
A typical situation for consumers may be, that the probability returned by the probability forecast is too low.
In such cases, the consumer may want to know how to set the intervals in order to make sure to find at least one matching provider for a specific SLA. Therefore we developed an optimization algorithm within our simulation. Within this paper we assume that a probability of greater than $99\%$ means that a match is guaranteed. For this case we use the term \textit{sure}.

Precondition for the optimization is to assign priorities to the service parameter of the requested SLA. A high priority means that the service parameter is very important for the consumer and that the interval should not be adapted in order to find at least one matching provider.

The optimizer starts extending the service parameter interval with the lowest priority. If this is not enough to reach a probability of $99\%$, the next service parameter with the second lowest parameter is extended. This algorithm is depicted in algorithm~\ref{alg:optimization}.

\begin{algorithm}
  \KwData{Requested SLA service parameters prioritized}
  \KwResult{Adapted service parameter intervals}
  sort list of services by priority\;
  set iterator to service with lowest priority\;
  	 	\While {true}{
   				calculate probability of finding at least one provider\;
    		  \If {probability>99\%} {    		  	  		 	  	
    		     break\;
   	 	  	  }
    		  \If {length==100} {
    		     next service\;
    		  }\Else{
    		  	service length++\;
    		  }
  }
\caption{Interval optimization}
\label{alg:optimization}
\end{algorithm}

In this section we further extend the example from section~\ref{subsec:exampleII}. The probability of finding at least one matching service was about $84\%$ and consequently lower than the limit of $99\%$. The priorities of the SLA are shown in table~\ref{tab:priorities}. 

\begin{table}[htbp]
\caption{Service priorities}
\begin{center}
\begin{tabular}{|l|c|c|}
\hline
\hline
\textbf{Service} & \textbf{Priority} & \textbf{Length} \\
\hline
\hline
Service A & 2 & 20\\
Service B & 1 (highest) & 30\\
Service C & 3 (lowest) & 84\\
\hline
\end{tabular}
\end{center}
\label{tab:priorities}
\end{table}


Algorithm~\ref{alg:optimization} calculates the results shown in the rightmost column of table~\ref{tab:priorities}. Service C, the service with the lowest priority, was extended to a length of $84$, the other services remained stable. With this server parameter lengths, the consumer can be sure that it finds at least one provider offering a matching SLA by consulting $20$ providers.


\section{Negotiation Range Analysis}
Until now we only checked, if the \textit{provider intervals} overlap with the \textit{consumer intervals}. We neglected the depth of overlapping. A deep intersection leads to a big negotiating range, whereas a small intersection leads to a small negotiating range.

To the best of the authors' knowledge no mathematical analysis of the negotiating depth is available. Therefore we executed a simulation in order to identify a distribution. Again, the \textit{provider interval's} length as well as position, are generated randomly. We executed the simulation with a \textit{consumer interval} length of $10$, $20$, $30$, $40$, $50$, $60$, $70$, $80$, $90$ and $100$. For each consumer length, we executed the simulation $1000000$ times and calculated the average negotiating range. The results of our simulation are shown in table~\ref{tab:overlappingDistance}.

\begin{table}[htbp]
\caption{Overlapping distance}
\begin{center}
\begin{tabular}{|l|c|c|c|c|c|}
\hline
\hline
interval length & 10 & 20 & 30 & 40 & 50  \\
negotiation range & 8,01 & 13,56 & 17,63 & 20,76 & 23,22 \\
\hline
interval length  & 60 & 70 & 80 & 90 & 100  \\
negotiation range  & 25 & 26,86 & 28,23 & 29,44 & 30,43   \\
\hline
\end{tabular}
\end{center}
\label{tab:overlappingDistance}
\end{table}

\begin{figure}[htb]
	\centering
		\includegraphics[width=0.6\linewidth]{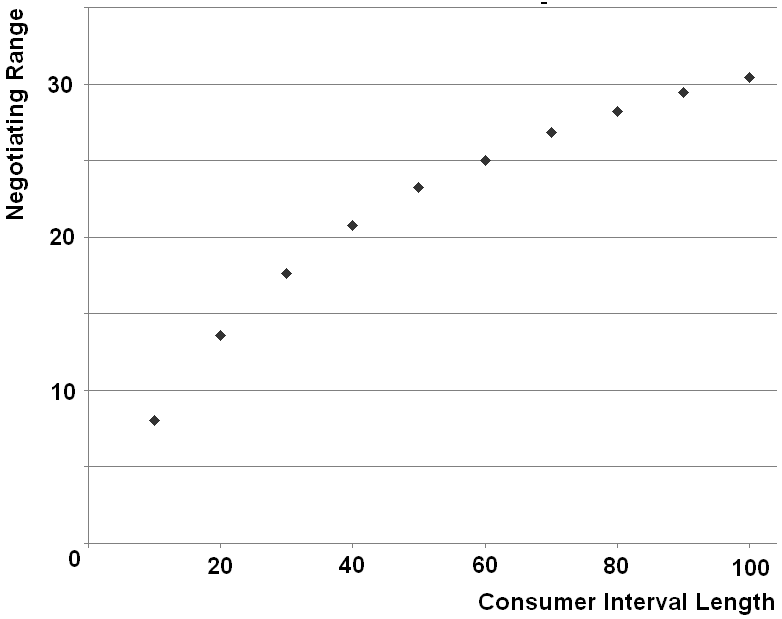}
		\caption{Negotiating range}
	\label{fig:negotiatingRangePointCloud}
	\end{figure}

In order to forecast the negotiation range of a SLA, containing any number of services with any length, we have to approximate the simulation results.
The results of table~\ref{tab:overlappingDistance} are visualized in figure~\ref{fig:negotiatingRangePointCloud}. It can be seen, that there is not a linear correlation between the negotiating range and the \textit{consumer interval} length. In this case, a linear regression trend line would lead to inappropriate forecasts. 

\begin{figure}[htb]
		\centering
		\subfigure[Scatter Plot]{\label{fig:sub:aII}\includegraphics[width=.4\linewidth]{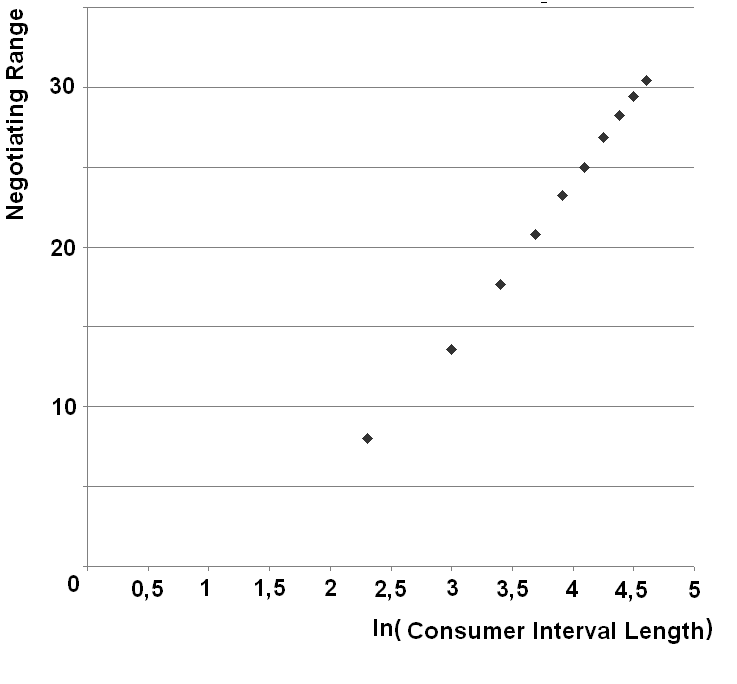}}\hfill
		\subfigure[Trend Line]{\label{fig:sub:bII}\includegraphics[width=.4\linewidth]{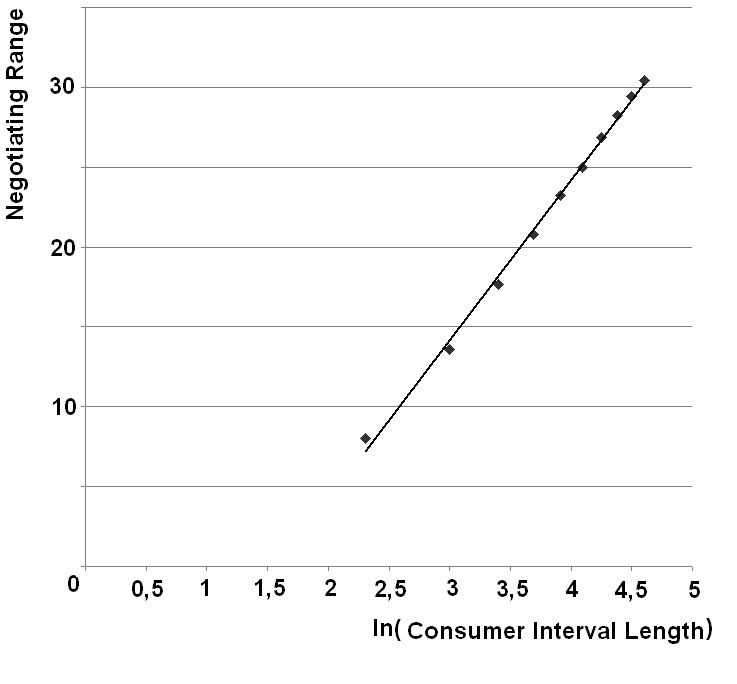}}
		\caption{Negotiation range transformed}
	\label{fig:NegotiationRangeLineareRegression}
	\end{figure}

By a logarithmic transformation $x_{new}=ln(x)$ of the data we can apply the linear regression trend line as approximation.
   The simulation after transformation is given in table~\ref{tab:overlappingDistanceLN} and visualized in figure~\ref{fig:NegotiationRangeLineareRegression}. Concluding from the scatter plot we apply linear regression to approximate the data analogously to section~\ref{subsec:linearRegression}.

\begin{table}[htbp]
\caption{Overlapping distance transformed}
\begin{center}
\begin{tabular}{|l|c|c|c|c|c|}
\hline
\hline
ln(interval length) & 2,303 & 2,996 & 3,401 & 3,689 & 3,912  \\
negotiation range & 8,01 & 13,56 & 17,63 & 20,76 & 23,22 \\
\hline
ln(interval length)  & 4,094 & 4,248 & 4,382 & 4,5 & 4,605  \\
negotiation range  & 25 & 26,86 & 28,23 & 29,44 & 30,43   \\
\hline
\end{tabular}
\end{center}
\label{tab:overlappingDistanceLN}
\end{table}

First of all, the averages of the two variables are calculated. Instead of the $x$ values we use the $ln$-transformed $x$ values.

\begin {equation}
\bar{X}=\frac{1}{n} \sum\limits_{i=1}^n x_{i}=3,813
\end{equation}

\begin {equation}
\bar{Y}=\frac{1}{n} \sum\limits_{i=1}^n y_{i}=22,314
\end{equation}

Then we calculate the slope.

\begin{equation}
s=\frac{\sum\limits_{i=1}^n ((x_{i}-\bar{X})
{(y_{i}-\bar{Y})})}{\sum\limits_{i=1}^n (x_{i}-\bar{X})^{2}}= 10,01
\end{equation}

Now we are able to transform the general linear equation illustrated below.

\begin{equation}
	\begin{aligned}
	& y-\bar{Y}=s \cdot (x-\bar{X}) \\
	& y-22,314=10,01 \cdot (x-3,813) \\
	& y-22,314=10,01 \cdot x-38,16813 \\
	& y=10,01 \cdot x- 15,85413\\
	\end{aligned}
	\label{equ:logarithmicFormula}
\end{equation}

After calculation of the trend line we have to check its quality by calculating its coefficient $SSE$.

\begin{equation}
	SSE=\sum\limits_{i=1}^n (y_{i_{observed}} - y_{i_{predicted}})^2=1,607
\end{equation}


\begin{equation}
	\begin{split}	
	R^2=1-\frac{SSE}{SS_{t}}=1-\frac{\sum\limits_{i=1}^n (y_{i_{observed}} - 	 y_{i_{predicted}})^2}{\sum\limits_{i=1}^n (y_{i_{observed}} - \bar{Y_{i}})^2} \\
	=\frac{\sigma_{y_{predicted}}^2}{\sigma_{y_{observed}}^2}
	\end{split}
\end{equation}

The variances are calculated by the equations below.

\begin{equation}
	\sigma_{y_{observed}}^2=\frac{\sum\limits_{i=1}^n (y_{i_{observed}}-
	\bar{Y})^2}{number of values} = 48,594164
\end{equation}

\begin{equation}
	\sigma_{y_{predicted}}^2=\frac{\sum\limits_{i=1}^n (y_{i_{predicted}}-
	\bar{Y})^2}{number of values} = 48,4336221
\end{equation}

Finally, we calculate the coefficient of determination.

\begin{equation}
	R^2=\frac{\sigma_{y_{predicted}}^2}{\sigma_{y_{observed}}^2} = 0,9967
\end{equation}

By the coefficient of determination near of $1$, the approximation can be considered as pretty good.

Due to the fact that we used a logarithmic transformation, we have to be careful making a prediction by the trend line, e.g. if a consumer wants to predict its negotiating range for one service of length $60$, a logarithmic transformation has to be applied by $x_{new}=ln(60)=4,094$.
The new $x$ value can be used in the trend line equation.

\begin{equation}
	y_{predicted}=10,01 \cdot 4,094-15,85413=25,127
\end{equation}

So the predicted negotiating range of about $25,1$ is in line to the simulated value $25$, as shown in table~\ref{tab:overlappingDistanceLN}.

\section{Model Use Case}
\label{sec:usecase}
In this section we want to provide a comprehensive example using all the approaches we developed before.
We assume that a consumer requests a SLA containing five services as shown in table~\ref{tab:slaServices}.

\begin{table}[htbp]
\caption{SLA services}
\begin{center}
\begin{tabular}{|l|c|c|}
\hline
\textbf{Service Name} & \textbf{Range} &  \textbf{Priority} \\
\hline
\hline
Service A & 20 & 1 (highest)\\
Service B & 30 & 2\\
Service C & 20 & 3\\
Service D & 70 & 4\\
Service E & 80 & 5 (lowest)\\
\hline
\end{tabular}
\end{center}
\label{tab:slaServices}
\end{table}

By this example we will answer the following questions:

\begin{itemize}
	\item What is the probability of finding at least one matching provider?
	\item What is the expected negotiating range?
	\item Which intervals should be adapted to be practically sure to find at least one matching provider?
	\item How many providers have to be consulted in order to be practically sure to find at least one matching provider?
\end{itemize}

First of all, we start calculating the single probability of finding a matching provider for each service parameter of the SLA by using the trend line equation (\ref{equ:linearRegressionFormular}):

\begin{equation}
	\begin{aligned}
	&P(\text{Service A})=0,00688667 \cdot 20 + 0,31133315=44,9\%\\
	&P(\text{Service B})=0,00688667 \cdot 30 + 0,31133315= 51,79\%\\
	&P(\text{Service C})=0,00688667 \cdot 20 + 0,31133315= 44,9\%\\
	&P(\text{Service D})=0,00688667 \cdot 70 + 0,31133315= 79,34\%\\
	&P(\text{Service E})=0,00688667 \cdot 80 + 0,31133315= 86,22\%\\
	\end{aligned}
	\label{equ:singlePropbabilitiesExample}
\end{equation}

 Now we combine the single probabilities by multiplying them as shown in equation (\ref{equ:probabilityMultiplicationRule}).

\begin{equation}
	\begin{aligned}
	& P_{\text{total probability}}= \\
	& P(\text{Service A}) \cdot P(\text{Service B}) \cdot P(\text{Service C}) \cdot \\
	&  P(\text{Service D}) \cdot P(\text{Service E})= \\
	& 0,449 \cdot 0,5179 \cdot 0,449 \cdot 0,7934 \cdot 0,8622=\\
	& 0,071423
	\end{aligned}
	\label{equ:connectedProbabilityExample}
\end{equation}

In order to get the probability of finding at least one matching provider consulting 20 providers, we have to use the binomial distribution as defined in equation (\ref{equ:binomialDistributionFormula}).

\begin{equation}
	\begin{aligned}
	&\left({{n}\atop {k}}\right)=\frac{n!}{(n-k)! \cdot k!}=\frac{20!}{(20-0)! \cdot 0!}=1 \\
	&P(k|p,n)=\left({{n}\atop {k}}\right) \cdot p^{k} \cdot (1-p)^{n-k}=\\
	&1 \cdot 0,071423^{0} \cdot (1-0,071423)^{20-0}=0,227
	\end{aligned}
\end{equation}

 The probability of finding one matching provider is about $22,7\%$. To get the probability of finding at least one matching provider we use the inverse probability equation (\ref{equ:inverseProability}).

\begin{equation}
	\begin{aligned}
	&P(\text{at least one success})=\\
	&1-P(\text{no success})=1-0,227=0,773
	\end{aligned}
\end{equation}

\begin{figure}[htb]
	\centering
		\includegraphics[width=0.8\linewidth]{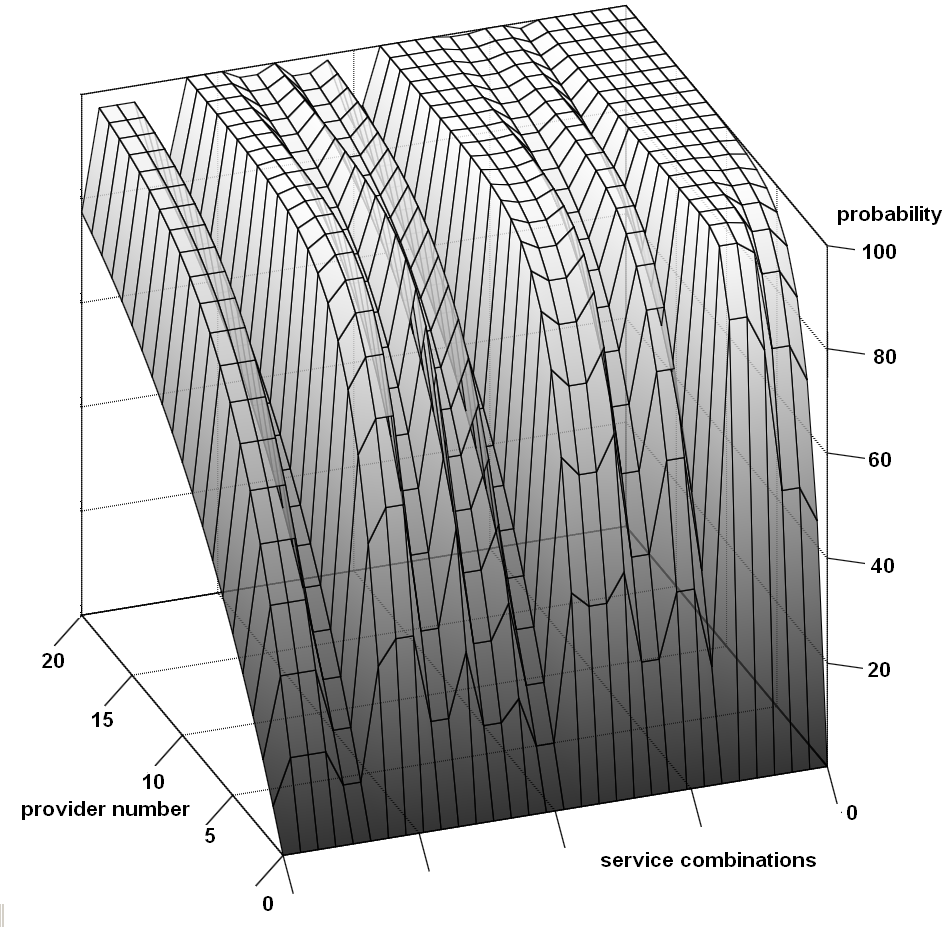}
		\caption{Probability for at least 1 matching provider}
	\label{fig:3dResult}
	\end{figure}

The probability of finding at least one matching provider is $77,3\%$.
Figure~\ref{fig:3dResult} illustrates the result as 3D plot. The ordinate shows the probability of finding at least one matching provider. The service combination axis represents all possible service combinations. The left most service combination is the combination containing all five services. The probability of finding a matching provider for these five services is $77,3\%$ consulting $20$ providers. The right most service combinations are the single services $A$, $B$, $C$, $D$ and $E$.

The number of combinations, which are represented on the ''service combinations'' axis in figure~\ref{fig:3dResult} can be calculated by combination without repetition~\cite{mazur2010}.

\begin{equation}
	\frac{n!}{(n-k)! \cdot k!}
\end{equation}

$n$ is the number of services and $k$ is the number of services.
%
We have to sum up the number of combinations using combinations consisting of $1$, $2$, $3$, $4$ and $5$ services. For $n=5$ it results in $31$ combinations.

\begin{equation}
	\sum\limits_{k=1}^{5} \frac{5!}{(5-k)! \cdot k!}=31
\end{equation}

Calculating the negotiating range we do a logarithmical transformation of the \textit{consumer intervals'} lengths ($x$ values) in order to use equation (\ref{equ:logarithmicFormula}).

\begin{equation}
	\begin{aligned}
	&ln(\text{Service A}_{length})= 2,996\\
	&ln(\text{Service B}_{length})= 3,401\\
	&ln(\text{Service C}_{length})= 2,996\\
	&ln(\text{Service D}_{length})= 4,248\\
	&ln(\text{Service E}_{length})= 4,382\\
	\end{aligned}
\end{equation}

Equation (\ref{equ:logarithmicFormula}) delivers the results shown in table \ref{tab:negotiationgRange}.
The sum of all negotiating ranges is $101,15$.

\begin{table}[htbp]
\caption{Negotiating ranges for use case}
\begin{center}
\begin{tabular}{|l|c|c|}
\hline
\textbf{Service Name} &  \textbf{Neg. Ranges} &  \textbf{Int. Opt.}\\
\hline
\hline
Service A &  14,14 & 20\\
Service B &  18,19 & 30\\
Service C &  14,14 & 84\\
Service D &  26,67 & 100\\
Service E &  28,01 & 100\\
\hline
\hline
$\sum$ &  101,15 & \\
\hline
\end{tabular}
\end{center}
\label{tab:negotiationgRange}
\end{table}

Finally, we adapt the service parameters in order to find a provider with a probability of at least $>99\%$. The priorities of the services are shown in table~\ref{tab:slaServices}. Executing the optimization algorithm~\ref{alg:optimization} leads to the optimized interval lengths shown in the rightmost column of table~\ref{tab:negotiationgRange}. Three services have to be adapted in order to find a provider with a probability greater than $99\%$ consulting $20$ providers.


If the consumer does not want to extend its service intervals, it has to increase the number of services consulted. Instead of consulting $20$ providers, the consumer has to consult $63$ providers in order to find at least one matching provider. Table~\ref{tab:numberOfProvidersExample} shows the probability of finding at least one matching provider depending on the providers consulted. This number can be calculated by an iterative approach. The number of providers is increased by one, until the probability of finding at least one provider exceeds $99\%$.

\begin{table}[htbp]
\caption{Probability depending on number of providers consulted}
\begin{center}
\begin{tabular}{|l|c|c|c|}
\hline
\textbf{Providers} &  \textbf{Probability} & \textbf{Providers} &  \textbf{Probability} \\
\hline
\hline
 20 & 77,3\% & 50 & 97,6\%\\
 25 & 84,4\% & 55 & 98,3\%\\
 30 & 89,2\% & 60 & 98,8\%\\
 35 & 92,5\% & 61 & 98,989\%\\
 40 & 94,9\% & 62 & 98,995\%\\
 45 & 96,5\% & 63 & 99,1\%\\
\hline
\end{tabular}
\end{center}
\label{tab:numberOfProvidersExample}
\end{table}

\section{Simulation Prototype}
\label{sec:prototype}

\begin{figure}[htb]
	\centering
		\includegraphics[width=0.9\linewidth]{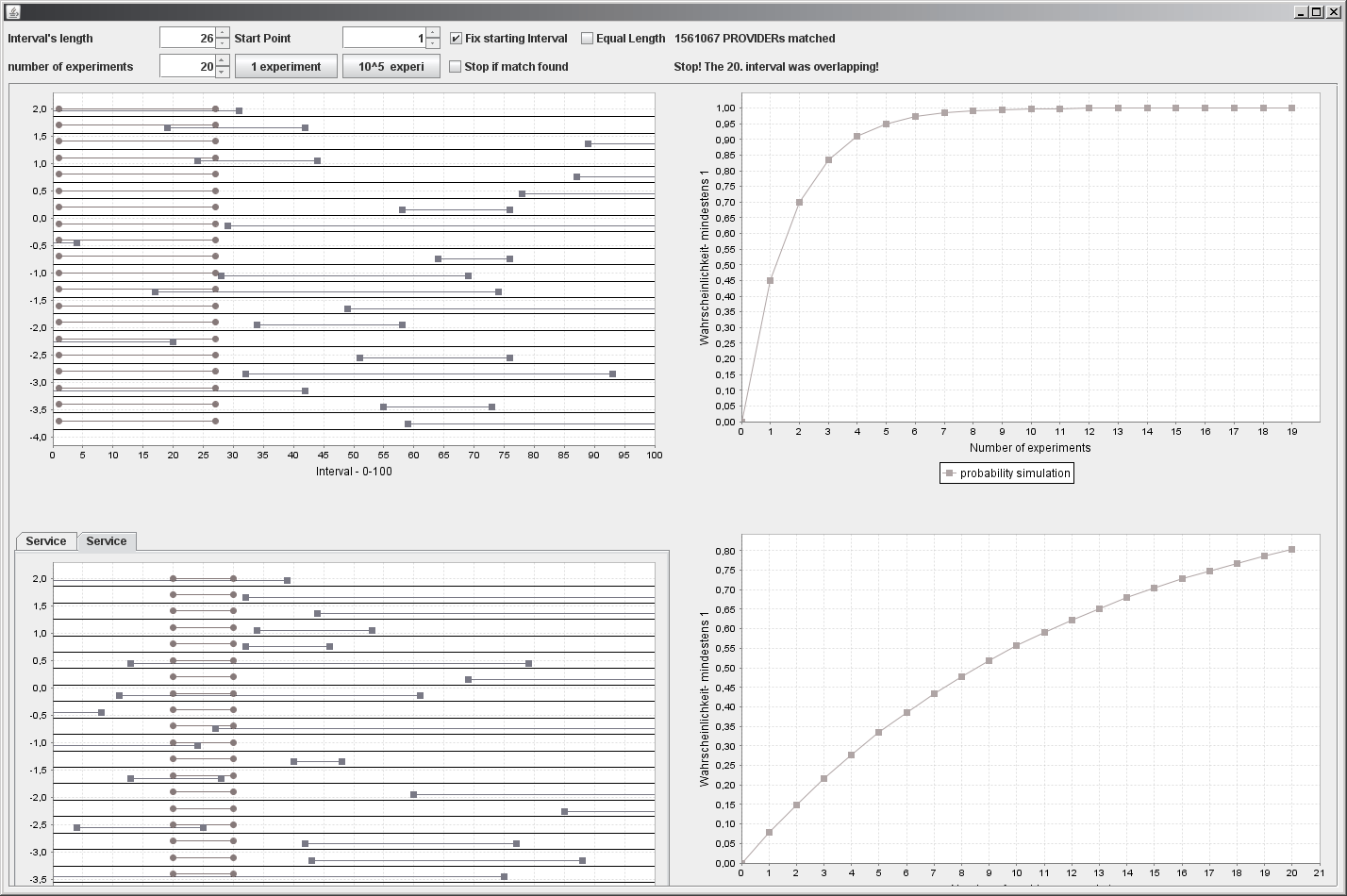}
		\caption{Simulator user interface}
	\label{fig:simulatorUserInterface}
	\end{figure}

We developed two prototypes in Java, the ''Simulator'' and the ''Cockpit''. The Simulator's user interface is depicted in figure~\ref{fig:simulatorUserInterface}.

\begin{figure}[htb]
	\centering
		\includegraphics[width=0.9\linewidth]{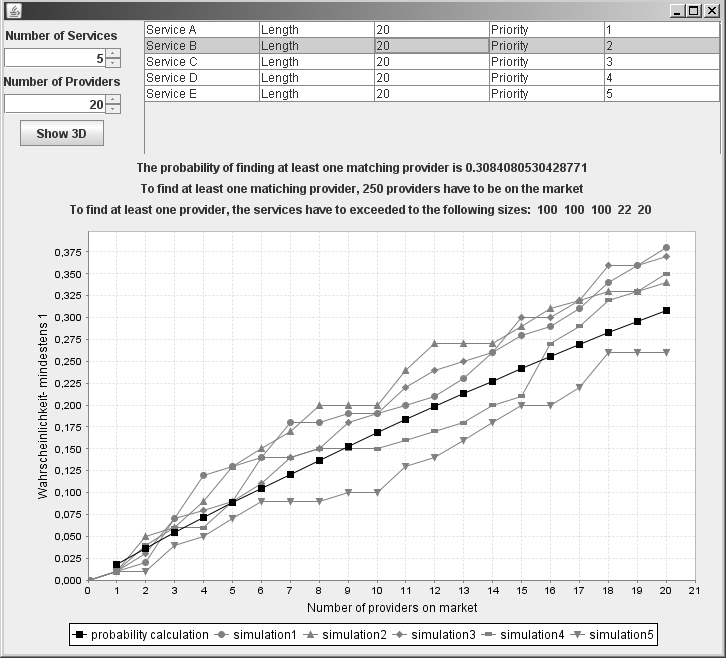}
		\caption{Cockpit user interface}
	\label{fig:cockpitUserInterface}
	\end{figure}

The Simulator reads the consumer's service parameter properties such as length and position from a XML file. These properties can be modified within the GUI.
Before starting simulation, the GUI allows to set the random \textit{provider interval} generation. The \textit{provider intervals} can be generated as described in this paper. Another option would be to generate the provider's intervals with a random position and fixed lengths.
The Simulator visualizes the simulation experiments by representing the consumer's and provider's ranges as intervals. Further the simulated probability distribution of finding at least one matching provider for a single service parameter as well as the probability distribution of finding at least one matching provider for a whole SLA is visualized. We used these graphics in figure~\ref{fig:probabilitySimulationScreenshot} and~\ref{fig:slaProbabilitySimulation}.
The goal of the second prototype Cockpit is to visualize the calculated forecasts. A screen shot of the Cockpit is given in figure~\ref{fig:cockpitUserInterface}.

This prototype allows to set the number of services, the number of providers and to modify the \textit{consumer interval} lengths as well as their priorities. 
The results are visualized as already shown in figure~\ref{fig:slaSimulationExampleII} and~\ref{fig:3dResult}.

\section{Conclusion}
\label{sec:conclusion}
In this paper we presented a comprehensive analytical model forecasting the chance of finding matching providers for web service negotiations based on quality of service parameters. The problem was to determine the probability of finding matching providers for a requested SLA. We simplified this problem by mapping it to the calculation of the probability that a given interval (\textit{consumer interval}) and a random generated interval (\textit{provider interval}) overlaps.  
Proved by simulation's results, we used regression and in further consequence probability calculation to do the forecasts. Thus, we are able to forecast the probability of finding matching providers for any number of services and any \textit{consumer interval} length.
Moreover, we introduced a simple mechanism adapting the ranges in order to increase the probability of finding one matching provider. Therefore, the consumer has to prioritize the service parameters.
Another way to increase the probability of finding a matching service, is to increase the number of providers consulted.
For justification we followed an approach checking our theoretical findings by simulation of practical examples.


This work is part of a research endeavour which aims for the development of a
framework for automatic, adaptive, and dynamic ne-gotiation and re-negotiation processes establishing an ICT marketplace for web services.
%
%

\end{document}